\documentclass{birkjour}
 \theoremstyle{definition}
 
 \theoremstyle{remark}

 \numberwithin{equation}{section}
 
\PassOptionsToPackage{original}{pict2e}
\usepackage{amsmath}
\usepackage{dsfont}
\usepackage{amsfonts}
\usepackage{amssymb}
\usepackage{color}
\usepackage{mathrsfs}
\usepackage{calc}
\usepackage{colonequals}
\usepackage{lmodern}
\usepackage[english]{babel}
\usepackage{xcolor}
\usepackage{upgreek}
\usepackage{babel,blindtext,scrpage2,eso-pic}
\usepackage{graphicx,pspicture}
\usepackage{colortbl}
\usepackage{mathtools}
\usepackage{array}
\usepackage{wrapfig}
\usepackage[makeroom]{cancel}
\usepackage{parcolumns}
\usepackage{framed}
\usepackage{booktabs}
\usepackage{psfrag}
\usepackage[margin=10pt,font=small,labelfont=bf,
labelsep=period]{caption}
\usepackage[bottom]{footmisc} 
\usepackage{indentfirst} 
\usepackage{upref} 
\usepackage{cite} 
\usepackage[rightcaption]{sidecap}
\usepackage{psfrag}
\usepackage{subfig}
\usepackage{float}
\usepackage{hyperref}

\newcommand{\crg}{{\rm CR}}
\newcommand{\nBound}{\mathcal{N}}
\newcommand{\scon}{\text{\rm sc}}

\makeatletter
\renewcommand*\env@matrix[1][*\c@MaxMatrixCols c]{%
  \hskip -\arraycolsep
  \let\@ifnextchar\new@ifnextchar
  \array{#1}}
\makeatother

\makeatletter
\def\hlinewd#1{%
\noalign{\ifnum0=`}\fi\hrule \@height #1 %
\futurelet\reserved@a\@xhline}
\makeatother

\renewcommand{\epsilon}{\varepsilon}

\newcommand{\cF}{\text{\rm Y}}

\renewcommand{\vec}[1]{\boldsymbol{#1}}
\newcommand{\gInst}{\mathring{g}}
\newcommand{\KkbarInst}{\mathring{\Kkbar}}

\newcommand{\cix}{I}

\makeatletter

\newcommand{\Rmnum}[1]{\expandafter\@slowromancap\romannumeral #1@}
\makeatother

\newcommand{\noR}{\mathds{R}}

\newcommand{\noC}{\mathds{C}}

\newcommand{\MaFs}{\mathcal{M}}  

\newcommand{\SR}{R}     
\newcommand{\Z}{D}      

\newcommand{\bg}{\bar{g}}   
\newcommand{\bZ}{\bar{\Z}}   

\newcommand{\cUV}{\mathscr{S}_{\text{UV}}}
 
\newcommand{\Tr}{\text{Tr}}   
\newcommand{\Order}[1]{\mathcal{O}(#1)}  

\newcommand{\md}{\text{d}}
 
\newcommand{\const}{\text{const}}    
 
\newcommand{\EAA}{\Gamma}  
 
\newcommand{\CosmConst}{\lambda}
\newcommand{\Kk}{\CosmConst}  
\newcommand{\Kkbar}{\Lambda}  

\newcommand{\KkbarD}{\Kkbar_k^{\dyn}}  
\newcommand{\KkbarB}{\Kkbar_k^{\background}}  

\newcommand{\background}{\text{B}}
\newcommand{\dyn}{\text{Dyn}}

\newcommand{\sm}{\text{sm}}

\newcommand{\KkD}{\CosmConst^{\dyn}} 

\newcommand{\IR}{{\rm IR} }
\newcommand{\UV}{{\rm UV} }
 
\newcommand{\tg}{g}

\newcommand{\nkD}{G_k^{\dyn}}
\newcommand{\nkbB}{G_k^{\background}}

\newcommand{\flcb}{h}

\newcommand{\cfunc}{\mathscr{C}}

\begin{document}

\title[Is there a $C$-function in 4D Quantum Einstein Gravity?]{Is there a $C$-function in\\ 4D Quantum Einstein Gravity?\footnote{Talk given by M.R. at {\it Quantum Mathematical Physics}, Regensburg, 2014.}}

\author[Becker]{Daniel Becker}

\address{%
Institute of Physics, University of Mainz\\
Staudingerweg 7, D-55099 Mainz, Germany}

\email{BeckerD@thep.physik.uni-mainz.de}

\thanks{}
\author[Reuter]{Martin Reuter}
\address{Institute of Physics, University of Mainz\\
Staudingerweg 7, D-55099 Mainz, Germany}
\email{Reuter@thep.physik.uni-mainz.de}


\date{\today}

\begin{abstract}
We describe a functional renormalization group-based method to search for `$C$-like' functions with properties similar to that in 2D conformal field theory.
It exploits the mode counting properties of the effective average action and is particularly suited for theories including quantized gravity.
The viability of the approach is demonstrated explicitly in a truncation of 4 dimensional Quantum Einstein Gravity, i.e. asymptotically safe metric gravity.
\end{abstract}

\maketitle

\section{Introduction}
A particularly intriguing result in 2-dimensional conformal field theory is  Zamolodchikov's $c$-theorem \cite{Zamol-c}.
It states that every 2D Euclidean quantum field theory with reflection positivity, rotational invariance, and a conserved energy momentum tensor possesses a function $C$ of its coupling constants, which is non-increasing along the renormalization group trajectories and is stationary at fixed  points where it equals the central charge of the corresponding conformal field theory.
After the advent of this theorem many authors tried to find a generalization that would be valid also in dimensions greater than two \cite{Cardy88,Osborn,Neto-Fradkin-93,Klebanov-Pufu-Safdi-11,Cappelli-Friedan-Latorre,Shore,Bastianelli,Anselmi98}.
This includes, for instance, suggestions by Cardy \cite{Cardy88} to integrate the trace anomaly of the energy-momentum tensor $\langle T_{\mu\nu} \rangle$ over a 4-sphere of unit radius, $C\propto \int_{S^4}\md^4 x\sqrt{g}\, \langle T^{\mu}_{\mu} \rangle$, the work of Osborn \cite{Osborn}, and ideas based on the similarity of $C$ to the thermodynamical free energy \cite{Neto-Fradkin-93}, leading to a conjectural `$F$-theorem' which states that, under certain conditions, the finite part of the free energy of 3-dimensional field theories on $S^3$ decreases along renormalization group (RG) trajectories and is stationary at criticality \cite{Klebanov-Pufu-Safdi-11}.
Cappelli, Friedan and Latorre \cite{Cappelli-Friedan-Latorre} proposed to define a $C$-function on the basis of the spectral representation of the 2-point function of the energy-momentum tensor.
While these investigations led to many important insights into the expected structure of the hypothetical higher-dimensional $C$-function, the search was successful only recently \cite{Komargo-Schwimmer,Lut-Pol-Rattaz} with the proof of the `$a$-theorem' \cite{Cardy88,Anselmi98}.
According to the $a$-theorem, the coefficient of the Euler form term in the induced gravity action of a 4D theory in a curved, but classical, background spacetime is non-increasing along RG-trajectories.
Clearly theorems of this type are extremely valuable as they provide non-perturbative information about quantum field theories or statistical systems in the strong coupling domain and constrain the structure of possible RG flows.

In this article  we are going to describe a functional RG-based search strategy by means of which `$C$-like' functions can possibly be identified  under rather general conditions, in particular in cases where the known $c$- and the $a$-theorems do not apply.
Our main motivation is in fact theories which include quantized gravity, in particular those based upon the Asymptotic Safety construction \cite{wein,mr,oliver1,oliver2,frank1,NJP,livrev,Blaubeuren}.

According to this strategy, the first step consists in trying to generalize the `counting property' of Zamolodchikov's $C$-function for a generic field theory in any number of dimensions: the sought-after function should roughly be equal to (or at least in a known way be related to) the number of degrees of freedom that are integrated out along the RG trajectory when the scale is lowered from the ultraviolet (UV) towards the infrared (IR).
Technically, we shall do this by introducing a higher-derivative mode-suppression factor in the underlying functional integral which acts as an IR cutoff.
We can then take advantage of the well established framework of the effective average action (EAA) to control the scale dependence \cite{wett-mr}, and to give a well defined meaning to the notion of a `number of modes'.

In a generic theory comprising a set of dynamical fields, $\Phi$, and associated background fields, $\bar{\Phi}$, the EAA is a `running' functional $\Gamma_k[\Phi,\bar{\Phi}]$  similar to the standard effective action, but with a built-in IR cutoff at a variable mass scale $k$.
Its $k$-dependence is governed by an exact functional RG equation (FRGE).
In this article we shall argue that there exists a natural and `essentially universal' map from $k$-dependent functionals $\Gamma_k$ to functions $\cfunc_k$ that are monotone along the flow and stationary at fixed points.
Here the term `universal' is to indicate that we would require only a few general properties to be satisfied, comparable to reflection positivity, rotational invariance, etc. in the case of Zamolodchikov's theorem.
The reason why we believe that there should exist such a map is that the respective monotonicity properties of $\Gamma_k$ and the $C$-function in 2D have essentially the same  simple origin.
They both `count' in a certain way the degrees of freedom (more precisely: fluctuation modes) that are already integrated out  at a given RG scale intermediate between the UV and the IR.

After a brief review of the necessary EAA apparatus, we shall present  a promising candidate for a quantity with properties close to a $C$-function.
It is obtained by evaluating $\Gamma_k[\Phi,\bar{\Phi}]$ at a particularly chosen pair of {\it $k$-dependent} arguments $(\Phi,\bar{\Phi})$,  namely $\Phi=\bar{\Phi}\equiv \bar{\Phi}_k^{\scon}$ where $\bar{\Phi}_k^{\scon}$ is a {\it self-consistent background field}.
By definition,  $\bar{\Phi}\equiv \bar{\Phi}_k^{\scon}$ is  self-consistent (`$\scon$') if the equation of motion for the dynamical field $\Phi$ derived from $\Gamma_k$ admits the solution $\Phi=\bar{\Phi}$.
With other words, if the system is put in a  background which  is  self-consistent, the fluctuations of the dynamical field, $\varphi\equiv \Phi-\bar{\Phi}$, have zero expectation value and, in this sense, do not modify this special background.
As we shall see, in theories without fermions,  $\cfunc_k\equiv\Gamma_k[\bar{\Phi}_k^{\scon},\bar{\Phi}_k^{\scon}]$ has indeed a number of attractive properties making it almost a $C$-function.
It is stationary at fixed points and it is  monotonically decreasing along the flow, at least when split-symmetry is broken only sufficiently weakly.

The latter restriction is crucial and requires an explanation.
In quantum gravity, Background Independence is a central requirement \cite{ARR} which, in
 the EAA framework, is met by employing the background field technique.
At the intermediate steps of the quantization one  introduces a background spacetime, equipped with a non-degenerate background metric in particular, but  makes sure that no observable prediction  depends on it.
This can be done by means of the Ward identities pertaining to the  split-symmetry \cite{gluco, elisa2, creh2} which governs the interrelation between $\varphi$ and $\bar{\Phi}$.
This symmetry, if intact, ensures that the physical contents of a theory is independent of the chosen background structures.
Usually, at the `off-shell' level of $\Gamma_k$, in particular when $k>0$, the symmetry is broken by the gauge fixing and cutoff terms in the bare action.
Insisting on unbroken split-symmetry in the physical sector restricts the admissible RG trajectories which the EAA may follow \cite{daniel2,daniel-eta}; only those which restore perfect split-symmetry at their end point ($k=0$) are acceptable.
The `sufficiently weak split-symmetry breaking' mentioned above is a related, but not exactly the same requirement, namely that the amount of symmetry breaking, on all scales $k\geq 0$, does not exceed a certain bound (given by eq. \eqref{eqn:cm_mot_043} below). 

Specifically we shall apply these ideas within the Asymptotic Safety approach to quantum gravity in the following \cite{wein,mr,oliver1,oliver2,frank1,NJP,livrev,Blaubeuren}.
The goal of the Asymptotic Safety program is to precisely define, and then to actually compute functional integrals over `all metrics' such as $\int \mathcal{D}\hat{g}\, e^{-S[\hat{g}_{\mu\nu}]}$.
The idea is to proceed indirectly and re-construct the integral from a solution of the FRGE for the EAA. 
Contrary to the functional integral, the FRGE is free from any UV singularities.
The nontrivial issue then consists in finding an RG trajectory consisting of {\it regular} action functionals $\{\Gamma_k[\,\cdot\,]\}$ that is {\it complete}, i.e. has well defined limits $k\rightarrow 0$ and $k\rightarrow\infty$, respectively.
Asymptotic Safety is a property that ensures the existence of the UV limit, $k\rightarrow\infty$.
Its prerequisite is a fixed point of the RG flow, $\Gamma_*$.
Let us assume there exists such a fixed point, and let $\cUV$ denote its UV critical manifold, that is the set of all actions which are pulled into $\Gamma_*$ under the inverse flow (i.e. when going from the IR to the UV).
Then, for the $k\rightarrow\infty$ limit to exist it is sufficient (and probably also necessary) to select any of the trajectories inside $\cUV$; we can then be sure that it has a singularity free UV behavior since it will always run into the fixed point at large scales and is easy to control then.

The only free choice in this entire construction concerns the {\it theory space}, $\mathcal{T}$, i.e. the space of functionals on which the FRGE operates; in particular the fields the functionals depend on, and their symmetries must be specified.
Given $\mathcal{T}$, the form of the FRGE and so ultimately also its fixed point properties are determined.
As $\Gamma_{k\rightarrow\infty}$ is closely related to the bare action $S$, we are actually {\it computing} $S$ from the fixed point condition, rather than putting it in `by hand'.
Knowing $\Gamma_*$ and the RG flow in its vicinity, and selecting an UV regularization scheme for the functional integral, one can in principle compute how the bare parameters on which this integral depends must be tuned in order to obtain a well defined limit when its UV regulator is removed, or the `continuum limit' is taken \cite{elisa1}.
For further details on Asymptotic Safety and the status of the program we refer to the reviews \cite{NJP,Blaubeuren,livrev}.

The rest of this article is organized as follows.
In section \ref{sec:02} we explain how the EAA can be used in order to `count' field modes, and we identify a natural candidate for a `$C$-function like' quantity that exists in any number of dimensions. 
In section \ref{sec:QEG} we apply these ideas to asymptotically safe metric gravity, or `Quantum Einstein Gravity' (QEG), and section \ref{sec:conc} contains the conclusions.

Our presentation follows ref. \cite{daniel-C} to which the reader is referred for additional details. 

\section{From the EAA to the \boldmath{$\cfunc$}-function} \label{sec:02}
We consider a  general quantum field theory on a $d$ dimensional Euclidean spacetime, either rigid or fluctuating, that is governed by a functional integral
$Z=\int \mathcal{D}\hat{\Phi}\, e^{- S[\hat{\Phi},\bar{\Phi}]}$.
The bare action $S$ depends on a set of commuting and anticommuting dynamical fields, $\hat{\Phi}$, and on a corresponding set of background fields, $\bar{\Phi}$.
In a  Yang-Mills theory, $\hat{\Phi}$ would contain both the gauge field and the Faddeev-Popov ghosts, and $S$ includes gauge fixing and ghost terms.
Furthermore, the corresponding background fields are part of $\bar{\Phi}$.
As a rule, the  {\it fluctuation field} $\hat{\varphi}\equiv \hat{\Phi}-\bar{\Phi}$ is always required to gauge-transform homogeneously, i.e. like a matter field.
Henceforth we regard $\hat{\varphi}$ rather than $\hat{\Phi}$ as the true dynamical variable and interpret $Z$ as an integral over the fluctuation variables: $Z=\int \mathcal{D}\hat{\varphi}\, \exp\left(-S[\hat{\varphi};\bar{\Phi}]\right)$.

The set of background fields, $\bar{\Phi}$, always contains a classical spacetime metric $\bg_{\mu\nu}$.
In typical particle physics applications on a rigid spacetime one is not interested in how $Z$ depends on this background metric and usually sets $\bg_{\mu\nu}=\delta_{\mu\nu}$ throughout.
Here in quantum gravity, where Background Independence is an issue, one needs to know $Z\equiv Z[\bg_{\mu\nu}]$ for {\it any} background.
In fact, employing the background field technique to implement Background Independence one represents  the dynamical metric as $\hat{g}_{\mu\nu}=\bg_{\mu\nu}+\hat{h}_{\mu\nu}$ and requires invariance under split-symmetry transformations $\left(\delta \bg_{\mu\nu}=-\epsilon_{\mu\nu},\,\delta \hat{h}_{\mu\nu}=\epsilon_{\mu\nu}\right)$ at the level of observable quantities \cite{daniel2}.
Assuming in the sequel that spacetime is dynamical, $\hat{g}_{\mu\nu}$ and $\hat{h}_{\mu\nu}$ are special  components of $\hat{\Phi}$ and $\hat{\varphi}$, respectively.

Picking a basis in field space, $\{\varphi_{\omega}\}$, we expand $\hat{\varphi}(x)=\sum_{\omega} a_{\omega}\, \varphi_{\omega}(x)$,
where $\sum_{\omega}$ stands for a summation and\slash or integration over all labels carried by the basis elements.
Then $\int \mathcal{D}\hat{\varphi}$ is interpreted as the integration over all possible values that can be assumed by the expansion coefficients $a\equiv \{a_{\omega}\}$.
Thus, $Z=\prod_{\omega} \int_{-\infty}^{\infty}\md a_{\omega}\, \exp\left(-S[a;\bar{\Phi}]\right)$.

Let us assume that the $\varphi_{\omega}$'s are eigenfunctions of a certain differential operator, $\mathcal{L}$, which may depend on the background fields $\bar{\Phi}$, and which has properties similar to the negative Laplace-Beltrami operator, $-\bZ^2$.
We suppose that $\mathcal{L}$ is built from covariant derivatives involving $\bg_{\mu\nu}$ and the background Yang-Mills fields, if any, so that it is covariant under spacetime diffeomorphism and gauge-transformations.
We assume an eigenvalue equation $\mathcal{L}\varphi_{\omega} =\Omega^2_{\omega} \varphi_{\omega}$ with positive spectral values $\Omega_{\omega}^2>0$.
The precise choice of $\mathcal{L}$ is arbitrary to a large extent.
The only property of $\mathcal{L}$ we need is that it should associate small (large) distances on the rigid spacetime equipped with the metric $\bg_{\mu\nu}$ to large (small) values of $\Omega_{\omega}^2$.
A first (but for us not the essential) consequence is that we can now easily install a UV cutoff by restricting the ill-defined infinite product $\prod_{\omega}$ to only those $\omega$'s  which satisfy $\Omega_{\omega}<\Omega_{\text{max}}$.
This implements a UV cutoff at the mass scale $\Omega_{\text{max}}$.

More importantly for our purposes, we also introduce a smooth IR cutoff at a variable scale $k\leq \Omega_{\text{max}}$ into the integral, replacing it with
\begin{align}
Z_k= {\prod_{\omega}}^{\prime} \int_{-\infty}^{\infty}\md a_{\omega}\, e^{-S[a;\bar{\Phi}]}e^{-\Delta S_k}
\label{eqn:cm_mot_001}
\end{align}
where the prime indicates the presence of the UV cutoff, and 
\begin{align}
\Delta S_k \equiv \frac{1}{2} \sum_{\omega} R_k(\Omega_k^{2})\, a_{\omega}^2
\label{eqn:cm_mot_002}
\end{align}
implements the IR cutoff.
The extra piece in the bare action, $\Delta S_k$, is designed in such a way that those $\varphi_{\omega}$-modes which have eigenvalues $\Omega_{\omega}^2 \ll k^2$ get suppressed by a small factor $e^{-\Delta S_k}\ll 1$ in eq. \eqref{eqn:cm_mot_001}, while $e^{-\Delta S_k}=1$ for the others.
The function $R_k$ is essentially arbitrary, except for its interpolating behavior between $R_k(\Omega_{\omega}^2)\sim k^2$ if $\Omega_{\omega} \ll k$ and $R_k(\Omega_{\omega}^2)=0$ if $\Omega_{\omega}\gg k$.

The operator $\mathcal{L}$ defines the precise notion of `coarse graining' field configurations.
We regard the $\varphi_{\omega}$'s with $\Omega_{\omega}>k$ as the `short wavelength' modes, to be integrated out first, and those with small eigenvalues $\Omega_{\omega}<k$ as the `long wavelength' ones whose impact on the fluctuation's dynamics is not yet taken into account.
This amounts to a diffeomorphism and gauge covariant generalization of the standard Wilsonian renormalization group, based on standard Fourier analysis on $\noR^d$, to situations with arbitrary background fields $\bar{\Phi}=(\bg_{\mu\nu},\bar{A}_{\mu},\cdots)$.

While helpful for the interpretation, it is often unnecessary to perform the expansion of $\hat{\varphi}(x)$ in terms of the $\mathcal{L}$-eigenfunctions explicitly.
Rather, one thinks of \eqref{eqn:cm_mot_001} as a `basis independent' functional integral 
\begin{align}
Z_k = \int \mathcal{D}^{\prime}\hat{\varphi}\, e^{-S[\hat{\varphi};\bar{\Phi}]}  e^{-\Delta S_k[\hat{\varphi};\bar{\Phi}]}
\label{eqn:cm_mot_003}
\end{align}
for which the eigen-basis of $\mathcal{L}$ plays no special role, while the operator $\mathcal{L}$ as such does so, of course.
In particular the cutoff action $\Delta S_k$ is now rewritten with $\Omega_{\omega}^2$ replaced by $\mathcal{L}$ in the argument of $R_k$:
\begin{align}
\Delta S_k[\hat{\varphi};\bar{\Phi}] = \frac{1}{2} \int \md^d x \sqrt{\bg}\, \hat{\varphi}(x)\, R_k(\mathcal{L})\, \hat{\varphi}(x)
\label{eqn:cm_mot_004}
\end{align}
Note that at least when $k>0$ the modified partition function $Z_k$ depends on the respective choices for $\mathcal{L}$ and $\bar{\Phi}$ separately.

The family of $k$-dependent partition functions $Z_k$ enjoys a simple property  which is strikingly reminiscent of the $C$-theorem in 2 dimensions.
Let us assume for simplicity that all component fields constituting $\hat{\varphi}$ are commuting, and that $\bar{\Phi}$ has been chosen $k$-independent.
Then \eqref{eqn:cm_mot_003} is a (regularized, and convergent for appropriate $S$) purely bosonic integral with a positive integrand which, thanks to the suppression factor $e^{-\Delta S_k}$, decreases with increasing $k$.
Therefore, $Z_k$ and the `entropy' $\ln Z_k$, are monotonically decreasing functions of the scale:
\begin{align}
\partial_k \ln Z_k <0
\label{eqn:cm_mot_005}
\end{align}
The interpretation of \eqref{eqn:cm_mot_005} is clear:
Proceeding from the UV to the IR by lowering the infrared cutoff scale, an increasing number of field modes get un-suppressed, thus contribute to the functional integral, and as a consequence the value of the partition function increases.
Thus, in a sense, $\ln Z_k$ `counts' the number of field modes that have been integrated out already.
Before we can make this intuitive argument more  precise  we must introduce a number of technical tools at this point.

{\bf\noindent Running actions.}
Introducing a source term for the fluctuation fields turns the partition functions $Z_k[J;\bar{\Phi}]\equiv e^{W_k[J;\bar{\Phi}]}$  into a generating functional:
\begin{align}
 e^{W_k[J;\bar{\Phi}]}= \int\!\! \mathcal{D}^{\prime}\hat{\varphi}\, \exp \left(-S[\hat{\varphi};\bar{\Phi}]  -\Delta S_k[\hat{\varphi};\bar{\Phi}]\! +\! \int\!\! \md^d x \sqrt{\bg}\, J(x)\hat{\varphi}(x)\right)
\label{eqn:cm_mot_006}
\end{align}
Hence the $\bar{\Phi}$- and $k$-dependent expectation value $\langle \hat{\varphi} \rangle\equiv \varphi$ reads
\begin{align}
\varphi(x)\equiv \langle \hat{\varphi}(x)\rangle=\frac{1}{\sqrt{\bg(x)}} \frac{\delta W_k[J;\bar{\Phi}]}{\delta J(x)}
\label{eqn:cm_mot_007}
\end{align}
If we can solve this relation for $J$ as a functional of $\bar{\Phi}$, the definition of the Effective Average Action (EAA), essentially the Legendre transform of $W_k$, may be written as
\begin{align}
\Gamma_k[\varphi;\bar{\Phi}] = \int\md^d x\sqrt{\bg}\, \varphi(x)J(x)-W_k[J;\bar{\Phi}] -\Delta S_k[\varphi;\bar{\Phi}]
\label{eqn:cm_mot_008}
\end{align}
with the solution to \eqref{eqn:cm_mot_007} inserted, $J\equiv J_k[\varphi;\bar{\Phi}]$.
In the general case, $\Gamma_k$ is the Legendre-Fenchel transform of $W_k$, with $\Delta S_k$ subtracted.

The EAA gives rise to a source-field relationship which includes an explicit cutoff term linear in the fluctuation field:
\begin{align}
\frac{1}{ \sqrt{\bg}}\frac{\delta \Gamma_k[\varphi;\bar{\Phi}]}{\delta \varphi(x)} +\mathcal{R}_k[\bar{\Phi}] \varphi(x)=J(x)
\label{eqn:cm_mot_009}
\end{align}
Here and in the following we write $\mathcal{R}_k\equiv R_k(\mathcal{L})$, and the notation $\mathcal{R}_k[\bar{\Phi}]$ is used occasionally to emphasize that the cutoff operator may depend on the background fields.
The solution to \eqref{eqn:cm_mot_009}, and more generally all fluctuation correlators $\langle \hat{\varphi}(x_1)\cdots \hat{\varphi}(x_n)\rangle$ obtained by multiple differentiation of $\Gamma_k$, are functionally dependent on the background, e.g. $\varphi(x)\equiv \varphi_k[J;\bar{\Phi}](x)$.
For the expectation value of the full, i.e. un-decomposed field $\hat{\Phi}=\bar{\Phi}+\hat{\varphi}$ we employ the notation 
$\Phi=\bar{\Phi}+\varphi$ with $\Phi\equiv \langle \hat{\Phi}\rangle$ and $ \varphi\equiv \langle \hat{\varphi}\rangle$.
Using the complete field $\Phi$ instead of $\varphi$ as the second independent variable, accompanying  $\bar{\Phi}$, entails the `bi-field' variant of the EAA,
\begin{align}
\Gamma_k[\Phi,\bar{\Phi}]\equiv \Gamma_k[\varphi; \bar{\Phi}]\big|_{\varphi=\Phi-\bar{\Phi}}
\label{eqn:cm_mot_011}
\end{align}
which, in particular, is always `bi-metric': $\Gamma_k[g_{\mu\nu},\cdots,\bg_{\mu\nu},\cdots]$.

Organizing the terms contributing to $\Gamma_k[\varphi;\bar{\Phi}]$ according to their {\it level}, i.e. their degree of homogeneity in the $\varphi$'s, we assume that
 the EAA admits a {\it level expansion} of the form $\Gamma_k[\varphi;\bar{\Phi}]=\sum_{p=0}^{\infty}\, \check{\Gamma}_k^{p}[\varphi;\bar{\Phi}]$
where $\check{\Gamma}_k^p[c\,\varphi;\bar{\Phi}]=c^p \, \check{\Gamma}_k^p [\varphi;\bar{\Phi}]$ for any $c>0$.

{\bf\noindent Self-consistent backgrounds.}
We are interested in how the dynamics of the fluctuations $\hat{\varphi}$ depends on the environment they are placed in, the background metric $\bg_{\mu\nu}$, for instance, and the other classical fields collected in $\bar{\Phi}$.
It would be instructive to know if there exist special backgrounds in which the fluctuations are particularly `tame' such that, for vanishing external source, they consists in at most small oscillations about a stable equilibrium, with a vanishing mean: $\varphi\equiv\langle \hat{\varphi}\rangle=0$.
Such distinguished backgrounds $\bar{\Phi}\equiv\bar{\Phi}^{\scon}$ are referred to as {\it self-consistent} ($\scon$) since, if we prepare the system in one of those, the expectation value of the field $\langle \hat{\Phi} \rangle=\Phi=\bar{\Phi}$ does not get changed by  violent $\hat{\varphi}$-excitations that, generically, can shift the point of equilibrium.
From eq. \eqref{eqn:cm_mot_009} we obtain the following condition $\bar{\Phi}^{\scon}$ must satisfy (since $J=0$ here by definition):
\begin{align}
\frac{\delta}{\delta \varphi(x)} \Gamma_k[\varphi;\bar{\Phi}]\big|_{\varphi=0, \bar{\Phi}=\bar{\Phi}_k^{\scon}}=0
\label{eqn:cm_mot_013}
\end{align}
This is the {\it tadpole equation} from which we can compute the self-consistent background configurations, if any.
In general $\bar{\Phi}^{\scon}\equiv \bar{\Phi}^{\scon}_k$ will have an explicit dependence on $k$.
A technically convenient feature of \eqref{eqn:cm_mot_013} is that it  no longer contains the somewhat disturbing $\mathcal{R}_k \varphi$-term that was present in the general field equation \eqref{eqn:cm_mot_009}.
Self-consistent backgrounds are equivalently characterized by eq. \eqref{eqn:cm_mot_007},
\begin{align}
\frac{\delta}{\delta J(x)} W_k[J;\bar{\Phi}]\big|_{J=0, \bar{\Phi}=\bar{\Phi}_k^{\scon}}=0
\label{eqn:cm_mot_014}
\end{align}
which again expresses the vanishing of the fluctuation's one-point function.
Note that provided the level expansion exists we may replace \eqref{eqn:cm_mot_013} with 
\begin{align}
\frac{\delta}{\delta \varphi(x)} \check{\Gamma}^1_k[\varphi;\bar{\Phi}]\big|_{\varphi=0, \bar{\Phi}=\bar{\Phi}_k^{\scon}}=0
\label{eqn:cm_mot_015}
\end{align}
which involves only the level-(1) functional $\check{\Gamma}^1_k$.
Later on in the applications this trivial observation has the important consequence that {\it self-consistent background field configurations $\bar{\Phi}^{\scon}_k(x)$ can contain only running coupling constants of level $p=1$}, that is, the couplings parameterizing the functional  $\check{\Gamma}_k^1$ which is linear in $\varphi$.

In our later discussions the value of the EAA at $\varphi=0$ will be of special interest.
While it is still a rather complicated functional for a generic background where $\Gamma_k[0;\bar{\Phi}]=-W_k[J_k[0;\bar{\Phi}];\bar{\Phi}]$, the source which is necessary to achieve $\varphi=0$ for self-consistent backgrounds is precisely $J=0$, implying
\begin{align}
\Gamma_k[0;\bar{\Phi}_k^{\scon}]\equiv \check{\Gamma}_k^0[0;\bar{\Phi}_k^{\scon}]=-W_k[0;\bar{\Phi}_k^{\scon}]
\label{eqn:cm_mot_016}
\end{align}
Here we also indicated that in a level expansion  only the $p=0$ term of $\Gamma_k$ survives putting $\varphi=0$.
{Note that $\Gamma_k[0;\bar{\Phi}_k^{\scon} ]$ can contain only couplings of the levels $p=0$ and $p=1$, respectively, the former entering via $\check{\Gamma}_k^0$, the latter via $\bar{\Phi}_k^{\scon}$.}

{\noindent\bf FIDE, FRGE, and WISS.}
The EAA satisfies a number of important exact functional equations which include a functional integro-differential equation (FIDE), the functional RG equation (FRGE),  the Ward identity for the Split-Symmetry (WISS), and the BRS-Ward identity.

In full generality, the FIDE reads
\begin{align*}
e^{-\Gamma_k[\varphi;\bar{\Phi}]} &= \int \mathcal{D}^{\prime}\hat{\varphi}\, \exp\left(-S[\hat{\varphi};\bar{\Phi}]-\Delta S_k[\hat{\varphi};\bar{\Phi}] + \int \md^d x\, \hat{\varphi}(x) \frac{\delta \Gamma_k}{\delta \varphi(x)}[{\varphi};\bar{\Phi}] \right)
\end{align*}
The last term on its RHS, the one linear in $\hat{\varphi}$, vanishes if the background is self-consistent and, in addition, $\varphi=0$ is inserted:
\begin{align}
\exp\left(-\Gamma_k[0;\bar{\Phi}_k^{\scon}]\right) = \int \mathcal{D}^{\prime}\hat{\varphi}\, \exp\left(-S[\hat{\varphi};\bar{\Phi}_k^{\scon}]-\Delta S_k[\hat{\varphi};\bar{\Phi}_k^{\scon}] \right)
\label{eqn:cm_mot_018}
\end{align}
We shall come back to this important identity soon.

Another exact relation satisfied by the EAA is the FRGE,
\begin{align}
k\partial_k \Gamma_k[\varphi;\bar{\Phi}]=\frac{1}{2}\text{STr}\left[\left(\Gamma_k^{(2)}[\varphi;\bar{\Phi}]+\mathcal{R}_k[\bar{\Phi}] \right)^{-1} k \partial_k \mathcal{R}_k[\bar{\Phi}]\right]
\label{eqn:cm_mot_019}
\end{align}
comprising the Hessian matrix of the fluctuation derivatives $\Gamma_k^{(2)}\equiv \delta^2 \Gamma_k \slash \delta \varphi^2$.
The supertrace `$\text{STr}$' in \eqref{eqn:cm_mot_019} provides the additional minus sign which is necessary for the $\varphi$-components with odd Grassmann parity, Faddeev-Popov ghosts and fermions.

The action $\Gamma_k[\Phi,\bar{\Phi}]$ satisfies the following exact functional equation which governs the `extra' background dependence which it has over and above the one which combines with the fluctuations to form the full field $\Phi\equiv\bar{\Phi}+\varphi$:
\begin{align}
\frac{\delta}{\delta \bar{\Phi}(x)} \Gamma_k[\Phi,\bar{\Phi}]=\frac{1}{2}\text{STr}\left[\left(\Gamma_k^{(2)}[\Phi,\bar{\Phi}]+\mathcal{R}_k[\bar{\Phi}] \right)^{-1} \frac{\delta}{\delta \bar{\Phi}(x)} S_{\text{tot}}^{(2)}[\Phi,\bar{\Phi}]\right]
\label{eqn:cm_mot_020}
\end{align}
Here $S_{\text{tot}}^{(2)}$ is the Hessian of $S_{\text{tot}}=S+\Delta S_k$ with respect to $\Phi$, where $S$ includes gauge fixing and ghost terms.
The equation \eqref{eqn:cm_mot_020} is the Ward identity induced by the split-symmetry transformations $\delta \varphi=\epsilon$, $\delta\bar{\Phi}=-\epsilon$, hence the abbreviation `WISS'.
First obtained in \cite{gluco} for Yang-Mills theory, extensive use has been made of \eqref{eqn:cm_mot_020} in quantum gravity \cite{elisa2} as a tool to assess the degree of split-symmetry breaking and the reliability of certain truncations \cite{creh2}.

{\noindent\bf Pointwise monotonicity.} 
From the  definition of the EAA by a Legendre transform it follows that, for all $\bar{\Phi}$, the sum $\Gamma_k+\Delta S_k$ is a convex functional of $\varphi$, and that $\Gamma_k^{(2)}+\mathcal{R}_k$ is a strictly positive definite operator which can be inverted at all scales $k\in(0,\infty)$.
Now let us suppose that the theory under consideration contains Grassmann-even fields only.
Then the supertrace in \eqref{eqn:cm_mot_019} amounts to the ordinary, and convergent trace of a positive operator so that the FRGE implies
\begin{align}
k \partial_k \Gamma_k[\varphi;\bar{\Phi}]\geq 0\quad \text{ at all fixed }\, \varphi,\, \bar{\Phi}.
\label{eqn:cm_mot_021}
\end{align}
Thus, at least in this class of distinguished theories the EAA, evaluated at any fixed pair of arguments $\varphi$ and $\bar{\Phi}$, is a monotonically increasing function of $k$.
With other words, {\it lowering $k$ from the UV towards the IR the value of $\Gamma_k[\varphi;\bar{\Phi}]$ decreases monotonically}.
We refer to this property as {\it pointwise monotonicity} in order to emphasize that it applies at all points of field space, $(\varphi,\bar{\Phi})$, separately.

In presence of fields with odd Grassmann parity, fermions and Faddeev-Popov ghosts, the RHS of the FRGE is no longer obviously non-negative.
However,  if the only Grassmann-odd fields are ghosts the pointwise monotonicity \eqref{eqn:cm_mot_021} can still be made a general property of the EAA, the reason being as follows.
At least when one implements the gauge fixing condition strictly, it cuts-out a certain subspace of  the space of fields $\hat{\Phi}$ to be integrated over, namely the gauge orbit space.
Hereby the integral over the ghosts represents the measure on this subspace, the Faddeev-Popov determinant.
The subspace and its geometrical structures are invariant under the RG flow, however.
Hence the EAA pertaining to the manifestly Grassmann-even integral {\it over the subspace} is of the kind considered above, and the argument implying \eqref{eqn:cm_mot_021} should therefore be valid again.
For a more detailed form of this argument we must refer to \cite{daniel-C}.
From now on we shall make the explicit assumption, however, that the sets $\Phi$ and $\bar{\Phi}$ do not contain fermions.

{\noindent\bf Monotonicity vs. stationarity.}
The EAA evaluated at fixed arguments shares the monotonicity property with a $C$-function.
However, $\Gamma_k[\varphi;\bar{\Phi}]$ is not stationary at fixed points.
In order to see why, and how to improve the situation, some care is needed concerning the interplay of dimensionful and dimensionless variables, to which we turn next.

{\bf\noindent (A)}
Let us assume that the space constituted by the functionals of $\varphi$ and $\bar{\Phi}$ admits a basis $\{I_{\alpha}\}$ so that we can expand the EAA as 
\begin{align}
\Gamma_k[\varphi;\bar{\Phi}]=\sum_{\alpha} \bar{u}_{\alpha}(k)\,I_{\alpha}[\varphi;\bar{\Phi}]
\label{eqn:cm_mot_022}
\end{align}
with dimensionful running coupling constants $\bar{u}\equiv (\bar{u}_{\alpha})$.
They obey a FRGE in component form, $k\partial_k \bar{u}_{\alpha}(k)=\bar{b}_{\alpha}(\bar{u}(k);k)$, whereby the functions $\bar{b}_{\alpha}$ are defined by the expansion  $\text{Tr}[\cdots]= \sum_{\alpha} \bar{b}_{\alpha}(\bar{u}(k);k)\, I_{\alpha}[\varphi;\bar{\Phi}]$.

{\bf\noindent (B)}
Denoting the canonical mass dimension\footnote{Our conventions are as follows. We use dimensionless coordinates, $[x^{\mu}]=0$. Then $[\md s^2]=-2$ implies that all components of the various metrics have $[\hat{g}_{\mu\nu}]=[\bg_{\mu\nu}]=[g_{\mu\nu}]=-2$, and likewise for the fluctuations: $[\hat{h}_{\mu\nu}]=[\flcb_{\mu\nu}]=-2$.} of the running couplings by $[\bar{u}_{\alpha}]\equiv d_{\alpha}$, their dimensionless counterparts are defined by 
$u_{\alpha}\equiv k^{-d_{\alpha}} \bar{u}_{\alpha}$.
In terms of the dimensionless couplings the expansion of $\Gamma_k$ reads
\begin{align}
\Gamma_k[\varphi;\bar{\Phi}]=\sum_{\alpha} u_{\alpha}(k) k^{d_{\alpha}} I_{\alpha}[\varphi;\bar{\Phi}]
\label{eqn:cm_mot_024}
\end{align}
Now observe that since $\Gamma_k$ is dimensionless the basis elements have dimensions $\left[I_{\alpha}[\varphi;\bar{\Phi}]\right]=-d_{\alpha}$.
Purely by dimensional analysis, this implies that\footnote{We use the notation $c^{[\varphi]}\varphi \equiv \{c^{[\varphi_i]}\varphi_i\}$ for the set in which each field is rescaled according to its individual canonical dimension.}
\begin{align}
I_{\alpha}[c^{[\varphi]} \varphi; c^{[\bar{\Phi}]}\bar{\Phi}]=c^{-d_{\alpha}} I_{\alpha}[\varphi;\bar{\Phi}] \quad \text{ for any constant $c>0$.}
\label{eqn:cm_mot_025}
\end{align}
This relation expresses the fact that the nontrivial dimension of $I_{\alpha}$ is entirely due to that of its field arguments; there are simply no other dimensionful quantities available after the $k$-dependence has been separated off.
Using \eqref{eqn:cm_mot_025} for $c=k^{-1}$ yields the dimensionless monomials
\begin{align}
k^{d_{\alpha}} I_{\alpha}[\varphi;\bar{\Phi}]&= I_{\alpha}[k^{-[\varphi]}\varphi;k^{-[\bar{\Phi}]}\bar{\Phi}]
\equiv I_{\alpha}[\tilde{\varphi};\tilde{\bar{\Phi}}]
\label{eqn:cm_mot_026}
\end{align}
Here we introduced the sets of dimensionless fields,
\begin{align}
\tilde{\varphi}(x)\equiv k^{-[\varphi]} \varphi(x),\,\quad \tilde{\bar{\Phi}}(x)\equiv k^{-[\bar{\Phi}]}\bar{\Phi}(x)
\label{eqn:cm_mot_027}
\end{align}
which include, for instance, the dimensionless metric and its fluctuations:
\begin{align}
\tilde{\flcb}_{\mu\nu}(x)\equiv k^2 \flcb_{\mu\nu}(x),\,\quad \tilde{\bg}_{\mu\nu}(x)\equiv k^2 \bg_{\mu\nu}(x)
\label{eqn:cm_mot_028}
\end{align}
Exploiting \eqref{eqn:cm_mot_026} in \eqref{eqn:cm_mot_024} we obtain the following representation of the EAA which is entirely in terms of dimensionless quantities\footnote{Here one should also switch from $k$ to the manifestly dimensionless `RG time' $t\equiv \ln (k)+\const$, but we shall not indicate this notationally.} now:
\begin{align}
\Gamma_k[\varphi;\bar{\Phi}]=\sum_{\alpha} u_{\alpha}(k)\, I_{\alpha}[\tilde{\varphi};\tilde{\bar{\Phi}}]
\,\equiv\, \mathcal{A}_k[\tilde{\varphi};\tilde{\bar{\Phi}}]
\label{eqn:cm_mot_029}
\end{align}
Alternatively, one might wish to make its $k$-dependence explicit, writing,
\begin{align}
\Gamma_k[\varphi;\bar{\Phi}]=\sum_{\alpha} u_{\alpha}(k)\, I_{\alpha}[k^{-[\varphi]}\varphi;k^{-[\bar{\Phi}]}\bar{\Phi}]
\label{eqn:cm_mot_030}
\end{align}
In the second equality of \eqref{eqn:cm_mot_029} we introduced the new functional $\mathcal{A}_k$ which, by definition, is numerically equal to $\Gamma_k$, but its natural arguments are the dimensionless fields $\tilde{\varphi}$ and $\tilde{\bar{\Phi}}$.
Hence the $k$-derivative of $\mathcal{A}_k[\tilde{\varphi},\tilde{\bar{\Phi}}]$ is to be performed at fixed $(\tilde{\varphi},\tilde{\bar{\Phi}})$, while the analogous derivative of $\Gamma_k[\varphi;\bar{\Phi}]$ refers to fixed dimensionful arguments:
\begin{subequations}
\begin{align}
k\partial_k \mathcal{A}_k[\tilde{\varphi};\tilde{\bar{\Phi}}]&=\sum_{\alpha} k\partial_k u_{\alpha}(k)\, I_{\alpha}[\tilde{\varphi};\tilde{\bar{\Phi}}] \label{eqn:cm_mot_031A}\\
k\partial_k \Gamma_k[\varphi;\bar{\Phi}]&=\sum_{\alpha} \big\{ k\partial_k u_{\alpha}(k) + d_{\alpha} u_{\alpha}(k) \big\}\,k^{d_{\alpha}}\, I_{\alpha}[\varphi;\bar{\Phi}] \label{eqn:cm_mot_031B}
\end{align}
\label{eqn:cm_mot_031}
\end{subequations}

{\bf\noindent (C)}
The {\it dimensionless} couplings $u\equiv (u_{\alpha})$ can serve as local coordinates on {\it theory space,}$\mathcal{T}$.
By definition, the `points' of $\mathcal{T}$ are functionals $\mathcal{A}$ depending on dimensionless arguments: $\mathcal{A}[\tilde{\varphi};\tilde{\bar{\Phi}}]=\sum_{\alpha} u_{\alpha} \, I_{\alpha}[\tilde{\varphi};\tilde{\bar{\Phi}}]$.
Geometrically speaking, RG trajectories are curves $k\mapsto \mathcal{A}_k=\sum_{\alpha} u_{\alpha}(k)\, I_{\alpha}\in \mathcal{T}$ that are everywhere tangent to 
\begin{align}
k\partial_k \mathcal{A}_k=\sum_{\alpha} \beta_{\alpha}(u(k)) \, I_{\alpha}
\label{eqn:cm_mot_032}
\end{align}
The functions $\beta_{\alpha}$, components of a vector field $\vec{\beta}$ on $\mathcal{T}$, are obtained by translating $k\partial_k \bar{u}_{\alpha}(k)=\bar{b}_{\alpha}(\bar{u}(k);k)$ into the dimensionless language.
This leads to the autonomous system of differential equations
\begin{align}
k\partial_k u_{\alpha}(k)\equiv \beta_{\alpha}(u(k))=-d_{\alpha} u_{\alpha}(k) + b_{\alpha}(u(k))
\label{eqn:cm_mot_033}
\end{align}
Here $b_{\alpha}$, contrary to its dimensionful precursor $\bar{b}_{\alpha}$, has no explicit $k$-dependence, thus defining an RG-time independent vector field, the `RG flow' $(\mathcal{T},\vec{\beta})$.

If the flow has a fixed point at some $u^*$, i.e. $\beta_{\alpha}(u^*)=0$, the `velocity' of any trajectory passing this point vanishes there, $k\partial_k u_{\alpha}=0$.
Hence by \eqref{eqn:cm_mot_032} the action $\mathcal{A}_k$ becomes stationary there, that is, its scale derivative vanishes pointwise,
\begin{align}
k\partial_k \mathcal{A}_k[\tilde{\varphi};\tilde{\bar{\Phi}}]=0 \,\quad \text{ for all fixed }\, \tilde{\varphi},\, \tilde{\bar{\Phi}}\,.
\label{eqn:cm_mot_034}
\end{align}
So the entire functional $\mathcal{A}_k$ approaches a limit, $\mathcal{A}_*=\sum_{\alpha} u_{\alpha}^*\, I_{\alpha}$.
The standard EAA instead keeps running even in the fixed point regime:
\begin{align}
\Gamma_k[\varphi;\bar{\Phi}]=\sum_{\alpha} u_{\alpha}^*\, k^{d_{\alpha}}\, I_{\alpha}[\varphi;\bar{\Phi}]\quad \text{ when }\quad u_{\alpha}(k)=u_{\alpha}^*\,.
\label{eqn:cm_mot_035}
\end{align}

{\bf\noindent (D)}
This brings us back to the `defect' of $\Gamma_k$ we wanted to repair:
While $\Gamma_k[\varphi;\bar{\Phi}]$ was explicitly seen to decrease monotonically along RG trajectories,  it does not come to a halt at fixed points in general.
The redefined functional $\mathcal{A}_k$, instead, approaches a finite limit $\mathcal{A}_*$ at fixed points, but is it monotone along trajectories?

Unfortunately this is not the case, and the culprit is quite obvious, namely the $d_{\alpha}u_{\alpha}$-terms present in the scale derivative of $\Gamma_k$, but absent for $\mathcal{A}_k$.
The positivity of the RHS of eq. \eqref{eqn:cm_mot_031B} does not imply the positivity of the RHS of eq. \eqref{eqn:cm_mot_031A}, and {\it there is no obvious structural reason for $k\partial_k \mathcal{A}_k[\tilde{\varphi};\tilde{\bar{\Phi}}]\geq 0$ at fixed $\tilde{\varphi}$, $\tilde{\bar{\Phi}}$.}
The best we can get is the following lower bound for the scale derivative: $
k\partial_k \mathcal{A}_k[\tilde{\varphi};\tilde{\bar{\Phi}}]\geq - \sum_{\alpha} d_{\alpha} u_{\alpha}(k) \, I_{\alpha}[\tilde{\varphi};\tilde{\bar{\Phi}}]
$.

{\noindent\bf The proposal.}
The complementary virtues of $\mathcal{A}_k$ and $\Gamma_k$ with respect to monotonicity along trajectories and stationarity at critical points suggest the following strategy for finding a $C$-type function with better properties:
Rather than considering the functionals pointwise, i.e. with fixed configurations of either the dimensionless or dimensionful fields inserted, one should evaluate them at {\it explicitly scale dependent arguments}:
$\cfunc_k \stackrel{?}{=} \Gamma_k[\varphi_k;\bar{\Phi}_k]\equiv\mathcal{A}_k[\tilde{\varphi}_k;\tilde{\bar{\Phi}}_k]$.
The hope is that 
$\varphi_k\equiv k^{[\varphi]}\tilde{\varphi}_k$, and  $\bar{\Phi}_k\equiv k^{[\bar{\Phi}]}\tilde{\bar{\Phi}}_k$
can be given a $k$-dependence which is intermediate between the two extreme cases $(\varphi,\bar{\Phi})=\const$ and  $(\tilde{\varphi},\tilde{\bar{\Phi}})=\const$, respectively, so as to preserve as much as possible of the monotonicity properties of $\Gamma_k$, while rendering $\cfunc_k$ stationary at fixed points of the RG flow.

The most promising candidate which we could find so far is
\begin{align}
\cfunc_k=\Gamma_k[0;\bar{\Phi}^{\scon}_k]=\mathcal{A}_k[0;\tilde{\bar{\Phi}}^{\scon}_k]
\label{eqn:cm_mot_039}
\end{align}
Here the fluctuation argument is set to zero, $\varphi_k\equiv 0$, and for the background we choose a self-consistent one, $\bar{\Phi}^{\scon}_k$, a solution to the tadpole equation \eqref{eqn:cm_mot_013}, or equivalently its dimensionless variant
\begin{align}
\frac{\delta}{\delta \tilde{\varphi}(x)} \mathcal{A}_k[\tilde{\varphi};\tilde{\bar{\Phi}}]\big|_{\tilde{\varphi}=0,\, \tilde{\bar{\Phi}}=\tilde{\bar{\Phi}}^{\scon}_k}=0
\label{eqn:cm_mot_040}
\end{align}

The function $k\mapsto \cfunc_k$ defined by eq. \eqref{eqn:cm_mot_039} has a number of interesting properties to which we turn next.

\noindent {\bf(i) Stationarity at critical points.}
When the RG trajectory approaches a fixed point, $\mathcal{A}_k[\tilde{\varphi};\tilde{\bar{\Phi}}]$ approaches $\mathcal{A}_*[\tilde{\varphi};\tilde{\bar{\Phi}}]$ pointwise.
Furthermore, the tadpole equation \eqref{eqn:cm_mot_040} becomes $(\delta \mathcal{A}_*\slash \delta \tilde{\varphi})[0;\tilde{\bar{\Phi}}_*]=0$.
It is $k$-independent, and so is its solution, $\tilde{\bar{\Phi}}_*$.
Thus $\cfunc_k$ approaches a well defined, finite constant:
\begin{align}
\cfunc_k \xrightarrow{\text{FP}}{}\cfunc_*=\mathcal{A}_*[0;\tilde{\bar{\Phi}}_*]
\label{eqn:cm_mot_041}
\end{align}
Of course we can write this number also as $\cfunc_*=\Gamma_k[0;k^{[\bar{\Phi}]}\tilde{\bar{\Phi}}_*]$ wherein the explicit and the implicit scale dependence of the EAA cancel exactly when a fixed point is approached.

\noindent {\bf(ii) Stationarity at classicality.}
In a classical regime (`$\crg$'), by definition, $\bar{b}_{\alpha}\rightarrow 0$, so that the {\it dimensionful} couplings stop running: $\bar{u}_{\alpha}(k)\rightarrow \bar{u}_{\alpha}^{\crg}=\const$.
Thus, by \eqref{eqn:cm_mot_022}, $\Gamma_k$ approaches $\Gamma_{\crg}=\sum_{\alpha} \bar{u}_{\alpha}^{\crg}\, I_{\alpha}$ pointwise.
Hence the dimensionful version of the tadpole equation, \eqref{eqn:cm_mot_013}, becomes $k$-independent, and the same is true for its solution, $\bar{\Phi}^{\scon}_{\crg}$.
So, when the RG trajectory approaches a classical regime, $\cfunc_k$ asymptotes a constant:
\begin{align}
\cfunc_{k} \xrightarrow{\crg}{} \cfunc_{\crg}=\Gamma_{\crg}[0;\bar{\Phi}_{\crg}^{\scon}]
\label{eqn:cm_mot_042}
\end{align}
Alternatively we can write $\cfunc_{\crg}=\mathcal{A}_k[0;k^{-[\bar{\Phi}]} \bar{\Phi}^{\scon}_{\crg}]$ where it is now the explicit and implicit $k$-dependence of $\mathcal{A}_k$ which cancel mutually.

We observe that there is a certain analogy between `criticality' and `classicality', in the sense that dimensionful and dimensionless couplings exchange their roles.
The difference is that the former situation is related to special {\it points} of theory space, while the latter concerns  extended {\it regions} in $\mathcal{T}$.
In those regions, $\mathcal{A}_k$ keeps moving as $\mathcal{A}_k[\,\cdot\,]=\sum_{\alpha} \bar{u}_{\alpha}^{\crg}\,k^{-d_{\alpha}}\, I_{\alpha}[\,\cdot\,]$.
Nevertheless it is natural, and of particular interest in quantum gravity, to apply a (putative) $C$-function not only to crossover trajectories in the usual sense which connect two fixed points, but also to {\it generalized crossover transitions} where one of the fixed points is replaced by a classical regime.

\noindent {\bf(iii) Monotonicity at exact split-symmetry.}
If split-symmetry is exact in the sense that $\Gamma_k[\varphi;\bar{\Phi}]$ depends on the single independent field variable $\bar{\Phi}+\varphi\equiv \Phi$ only, and the theory is one of those for which pointwise monotonicity \eqref{eqn:cm_mot_021} holds true, then $k\mapsto \cfunc_k$ is a monotonically increasing function of $k$.
In fact, differentiating \eqref{eqn:cm_mot_039} and using the chain rule yields
\begin{align}
\partial_k \cfunc_k = \left(\partial_k \Gamma_k\right)[0;\bar{\Phi}_k^{\scon}] + \int\md^d x\left(\partial_k \bar{\Phi}_k^{\scon}(x)\right)\left. \left(\frac{\delta\Gamma_k }{\delta   \bar{\Phi}(x)} - \frac{\delta  \Gamma_k}{\delta \varphi(x)}\right)\right|_{\varphi=0,\, \bar{\Phi}=\bar{\Phi}_k^{\scon}}\!\!\!\!\!\!\!\!\!\!\!\!\!
\label{eqn:cm_mot_043}
\end{align}
In the first term on the RHS of \eqref{eqn:cm_mot_043} the derivative $\partial_k$ hits only the explicit $k$-dependence of the EAA.
By eq. \eqref{eqn:cm_mot_021} we know that this contribution is non-negative.
The last term, the $\delta\slash \delta \varphi$-derivative, is actually zero by the tadpole equation \eqref{eqn:cm_mot_013}.
Including it here it becomes manifest that the integral term in \eqref{eqn:cm_mot_043} vanishes when $\Gamma_k$ depends on $\varphi$ and $\bar{\Phi}$ only via the combination $\varphi+\bar{\Phi}$.
Thus we have shown that
\begin{align}
\partial_k \cfunc_k \geq 0 \,\quad \text{ at exact split-symmetry}
\label{eqn:cm_mot_044}
\end{align}

This is already close to what one should prove in order to establish $\cfunc_k$ as a `$C$-function'.
In particular in theories that require no breaking of split-symmetry the integral term in \eqref{eqn:cm_mot_043} is identically zero and we know that $\partial_k \cfunc_k\geq 0$ holds true.

Whether or not $\partial_k \cfunc_k$ is really non-negative for all $k$ depends on the size of the split-symmetry breaking the EAA suffers from.
To prove monotonicity of $\cfunc_k$ one would have to show on a case-by-case basis that the second term on the RHS of \eqref{eqn:cm_mot_043} never can override the first one, known to be non-negative, so as to render their sum negative.
In the next section we shall perform this analysis in a truncation of  Quantum Einstein Gravity, but  by working directly with the definition  $\cfunc_k=\Gamma_k[0;\bar{\Phi}_k^{\scon}]$ instead of eq. \eqref{eqn:cm_mot_043}.

{\noindent\bf Relating \boldmath{$\cfunc_k$} to a spectral density.}
Under special conditions, the EAA can be shown to literally `count' field modes.
For a sharp cutoff, and if $\mathcal{L}\equiv \Gamma_k^{(2)}[0;\bar{\Phi}_k^{\scon}]$ is such that it can be used as the cutoff operator, a formal calculation based upon the exact FRGE yields for the scale derivative of our candidate $\cfunc$-function:
\begin{align}
\frac{\md}{\md k^2} \cfunc_k = \Tr\left[\delta\left(k^2-\Gamma^{(2)}_k[0;\bar{\Phi}_k^{\scon}]\right)\right] \geq 0
\label{eqn:cm_mot_049}
\end{align}
This is exactly the spectral density of the Hessian operator, for the $\scon$-background and  vanishing fluctuations, a manifestly non-decreasing function of $k$.
If the $k$-dependence of $\Gamma_k^{(2)}$ is negligible relative to $k^2$, eq. \eqref{eqn:cm_mot_049} is easily integrated:
\begin{align}
\cfunc_k=\Tr\left[\Theta\left(k^2-\Gamma_k^{(2)}[0;\bar{\Phi}_k^{\scon}]\right)\right]+\text{const}
\label{eqn:cm_mot_050}
\end{align}
Thus, at least under the special conditions described and when the spectrum is discrete,  $\cfunc_k$ indeed counts field modes in the literal sense of the word.

Regardless of the present approximation we define in general
\begin{align}
\nBound_{k_1,k_2}\equiv \cfunc_{k_2}-\cfunc_{k_1}
\label{eqn:cm_mot_051}
\end{align}
Then, in the cases when the above assumptions apply and \eqref{eqn:cm_mot_050} is valid, $\nBound_{k_1,k_2}$ has a simple interpretation: it equals the number of eigenvalues between $k_1^2$ and $k_2^2>k_1^2$ of the Hessian operator $\EAA_k^{(2)}[0;\bar{\Phi}_k^{\scon}]$.
When the assumptions leading to \eqref{eqn:cm_mot_050} are not satisfied, the interpretation of $\nBound_{k_1,k_2}$, and $\cfunc_k$ in the first place, is less intuitive, but these quantities are well defined nevertheless.

\section{Asymptotically safe quantum gravity}\label{sec:QEG}
Next, we test the above $\cfunc_k$-candidate and apply it to  Quantum Einstein Gravity, a theory which is asymptotically safe most probably, that is, all physically relevant RG trajectories start out in the UV, for $k\,\text{`}\!=\!\!\text{'}\,\infty$, at a point  infinitesimally close to a non-Gaussian fixed point of the flow generated by the FRGE \eqref{eqn:cm_mot_019}.
When $k$ is lowered, the trajectories run towards the IR, always staying within the fixed point's UV critical manifold, and ultimately approach the (dimensionless) ordinary effective action.

Dealing with pure metric gravity here we identify $\Phi\equiv(g_{\mu\nu},\cdots)$, $\bar{\Phi}\equiv (\bg_{\mu\nu},\cdots)$, and $\varphi\equiv(\flcb_{\mu\nu},\cdots)$ as the dynamical, background, and fluctuation fields, respectively, where the dots stand for the entries due to the Faddeev-Popov ghosts.
To make the analysis technically feasible we are going to truncate the corresponding theory space.
Following ref. \cite{daniel-C} we focus here on the so-called bi-metric Einstein-Hilbert truncation.
The corresponding ansatz for the EAA has the structure $\EAA_k=\EAA_k^{\text{grav}}[g,\bg]+\cdots$ where the dots represent gauge fixing and ghost terms which are taken to be $k$-independent and of classical form.
The diffeomorphically invariant part of the action, $\Gamma_k^{\text{grav}}$,  comprises two separate Einstein-Hilbert terms built from the dynamical metric, $g_{\mu\nu}$, and its background analog, $\bg_{\mu\nu}$, respectively:
\begin{align}
  \EAA_k^{\text{grav}}[g,\bg]&= - \frac{1}{16\pi \nkD } \int\md^d x \sqrt{g}\, \big(\SR(g) - 2 \KkbarD\big)\nonumber  \\
&\quad   - \frac{1}{16\pi \nkbB }  \int\md^d x \sqrt{\bg}\, \big(\SR(\bg)- 2 \KkbarB\big) \label{eqn:trA07}
\end{align}
The 4 couplings $\left(\nkD,\,\KkbarD,\,\nkbB,\,\KkbarB\right)$ represent $k$-dependent  generalizations of the classical Newton and cosmological constant in the dynamical (`$\dyn$') and the background (`$\background$') sector, respectively.
In the simpler `single-metric' variant of the Einstein-Hilbert truncation \cite{mr} the difference between $g_{\mu\nu}$ and $\bg_{\mu\nu}$ is not resolved, and only one Einstein-Hilbert term is retained in $\EAA_k^{\text{grav}}$.
(Only in the gauge fixing term the two metrics appear independently.)

Expanding eq. \eqref{eqn:trA07} in powers of the fluctuation field $\flcb_{\mu\nu}=g_{\mu\nu}-\bg_{\mu\nu}$ yields the level-expansion of the EAA:
\begin{align}
  \EAA_k^{\text{grav}}[\flcb;\bg]&= - \frac{1}{16\pi G_k^{(0)} }  \int\md^d x \sqrt{\bg} \left(\SR(\bg) - 2\Kkbar_k^{(0)}\right)  \nonumber  \\
&\quad - \frac{1}{16\pi G_k^{(1)} }  \int\md^d x \sqrt{\bg}\, \Big[-\bar{G}^{\mu\nu}-\Kkbar_k^{(1)} \bg^{\mu\nu}\Big] \flcb_{\mu\nu} 
+ \Order{\flcb^2} \label{eqn:trA08}
\end{align}
In the level-description, the background  and  dynamical  couplings appear in certain combinations in front of  invariants that have a definite level, i.e. order in $\flcb_{\mu\nu}$.
The two sets of coupling constants are related by $1\slash G_k^{(0)}=1\slash G_k^{\background}+1\slash G_k^{\dyn}$ at level zero, and $G_k^{(p)}=G_k^{\dyn}$ at all higher levels $ p\geq1$, and similarly for the $\Lambda$'s.
Thus, by hypothesis, all couplings of level $p\geq1$ are assumed equal in this truncation.
In either parametrization the truncated theory space is 4-dimensional.

\begin{figure}[h!]
\centering
 \psfrag{a}[bc]{${\scriptstyle\text{\bf Type \Rmnum{3}a} }$}
  \psfrag{b}[m]{${\scriptstyle\text{\bf Type \Rmnum{1}a} }$}
 \psfrag{c}[tr]{${\scriptstyle\text{\bf Type \Rmnum{2}a} }$}
 \psfrag{g}{${ \tg }$}
  \psfrag{l}{${ \Kk }$}
\includegraphics[width=0.5\textwidth]{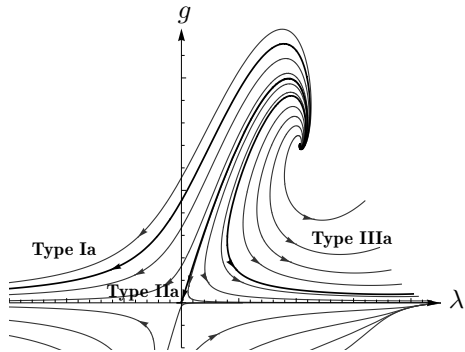}
 \caption{Phase portrait on the $\tg^{\dyn}$-$\KkD$ plane as obtained by projecting the 4-dimensional bi-metric flow.
 This projection is  qualitatively identical with the corresponding single-metric RG flow, displaying in particular the projection of a 4-dimensional non-Gaussian fixed point (NGFP).
}\label{fig:typeStructure}
\end{figure}

The beta-functions describing the flow of the dimensionless couplings $\tg_k^{\cix}\equiv k^{d-2}G_k^{\cix}$ and $\Kk_k^{\cix}\equiv k^{-2}\Kkbar_k^{\cix}$ for $\cix \in\{\background,\dyn,(0),(1)\}$ were derived and analyzed in \cite{mr,MRS2,daniel2}.
They were shown to give rise to both a trivial and a non-Gaussian fixed point (NGFP).
A 2-dimensional projection of the RG flow onto the $\tg^{\dyn}$-$\KkD$-plane is shown in Fig. \ref{fig:typeStructure}.
It is strikingly similar to the well known phase portrait of the corresponding single-metric truncation \cite{frank1}.
In this projection we can identify the same familiar classes of trajectories, namely those of 
type \Rmnum{1}a,  \Rmnum{2}a, or  \Rmnum{3}a, depending on whether the cosmological constant approaches $-\infty$, $0$, or $+\infty$ in the IR.
The type \Rmnum{3}a trajectories display a generalized crossover transition which connects a fixed point in the UV to a classical regime in the IR.
The latter is located on the trajectory's lower, almost horizontal branch where $\tg,\,\Kk\ll 1$ \cite{h3}.

{\noindent\bf Gravitational instantons.}
For the bi-metric Einstein-Hilbert truncation, the tadpole equation boils down to 
\begin{align}
R_{\mu\nu}(\bg_k^{\scon})=\frac{2}{d-2}\Kkbar_k^{(1)}\,\bg_{k\,\mu\nu}^{\scon}
\end{align}
so that the self-consistent backgrounds are Einstein spaces, $\mathcal{M}$, with  cosmological constant $\Kkbar_k^{(1)}$.
Furthermore, for $\cfunc_k$ to be finite, the manifold $\mathcal{M}$ must have a finite volume.
Trying to find backgrounds that exist for all scales  the simplest situation arises when all metrics $\bg_k^{\scon}$, $k\in [0,\infty)$ can be put on {\it the same smooth manifold} $\MaFs$, leading in particular to the same spacetime topology at all scales, thus avoiding the delicate issue of a topological change.
This situation is realized, for example, if the level-(1) cosmological constant is positive on all scales, as it is indeed the case along the type (\Rmnum{3}a) trajectories: $\Kkbar_k^{(1)}>0$, $k\in[0,\infty)$.

In the following we focus on this case, and we also specialize for $d=4$.
The requirement of a finite  volume is then met by a well studied class of Einstein spaces which exist for an arbitrary positive value of the cosmological constant, namely certain 4-dimensional gravitational instantons, such as Euclidean de Sitter space, $S^4$, the Page metric, the product space $S^2\times S^2$, or the Fubini-Study metric on the projective space $P_2(\noC)$ \cite{EGH}.
If $\gInst_{\mu\nu}$ is one of these instanton metrics for some reference value of the cosmological constant, $\KkbarInst$, simple scaling arguments imply that $\bg_{k\,\mu\nu}^{\scon}=\left(\KkbarInst\slash\Kkbar_k^{(1)}\right)\gInst_{\mu\nu}$ is a solution to the tadpole equation at any scale $k$.
Inserting it into the truncation ansatz for $\Gamma_k$ we find that  the function $k\mapsto \cfunc_k$ has the general structure
\begin{align}
\cfunc_k= \cfunc(\tg_k^{(0)},\Kk_k^{(0)},\Kk_k^{(1)})=\cF(g_k^{(0)}, \Kk_k^{(0)},\Kk_k^{(1)})\, \mathcal{V}(\MaFs,\gInst)
\label{eqn:intro_eq09}
\end{align}
Herein $\cF(\,\cdot\,)\equiv \cfunc(\,\cdot\,)\slash \mathcal{V}$ is given by the following function over  theory space:
\begin{align}
\cF(g^{(0)}, \Kk^{(0)},\Kk^{(1)})=- \frac{2\Kk^{(1)}-\Kk^{(0)}}{g^{(0)}\, (\Kk^{(1)})^2} \qquad \qquad \qquad \text{ $(d=4)$}
\label{eqn:intro_eq11}
\end{align}
Note that $\cfunc$ depends on both the RG trajectory and on the specific solution to the running self-consistency condition that has been picked, along this very trajectory. 
In eq. \eqref{eqn:intro_eq09} those two  dependencies factorize: the former enters via $\cF$, the latter via the dimensionless constant $\mathcal{V}(\MaFs,\gInst)\equiv \frac{1}{8\pi}\KkbarInst^2 \text{Vol}(\mathcal{M},\gInst)$. 
It characterizes the type of the gravitational instanton and can be shown to be actually independent of $\KkbarInst$.
For $S^4$, for instance, its value is $3\pi$, while the Fubini-Study metric has $\frac{9\pi}{2}$.
The dependence on the trajectory, parametrized as $k\mapsto \big(g_k^{(0,1)},\Kk_k^{(0,1)}\big)$, is obtained by evaluating a {\it scalar function on theory space} along this curve, namely $\cF:\mathcal{T}\rightarrow \noR$, $\big(g^{(0,1)},\Kk^{(0,1)}\big)  \mapsto \cF(g^{(0)}, \Kk^{(0)},\Kk^{(1)})$.
It is defined at all points of $\mathcal{T}$ where $\tg^{(0)}\neq 0$ and $\Kk^{(1)}\neq 0$, and turns out to be actually independent of $\tg^{(1)}$.

We shall refer to $\cF_k\equiv\cF(\tg_k^{(0)},\,\Kk_k^{(0)},\,\Kk_k^{(1)} )\equiv \cfunc_k\slash \mathcal{V}(\MaFs,\gInst)$ and $\cF(\,\cdot\,)\equiv \cfunc(\,\cdot\,)\slash \mathcal{V}(\MaFs,\gInst)$ as the {\it reduced $\cfunc_k$ and $\cfunc(\,\cdot\,)$ functions,} respectively.

{\noindent\bf Numerical results.}
In \cite{daniel2} the type \Rmnum{3}a trajectories on  $\left(\tg^{(0)},\,\Kk^{(0)},\,\tg^{(1)},\,\Kk^{(1)}\right)$-theory  space were analyzed in detail.
In \cite{daniel-C} representative examples were computed numerically, and then $\cfunc_k$ was evaluated along these trajectories.
Concerning the monotonicity of $\cfunc_k$, the results can be summarized as follows.

The set of RG trajectories that are asymptotically safe, i.e. originate in the UV at (or, more precisely, infinitesimally close to) the NGFP consists of two fundamentally different classes, namely those that are `physical' and restore split-symmetry at their end point $k=0$, and those which do not.
(Within the present truncation, and according to the lowest order of the WISS, eq. \eqref{eqn:cm_mot_020}, intact split-symmetry amounts to $\tg_k^{(0)}=\tg_k^{(1)}$ and $\Kk_k^{(0)}=\Kk_k^{(1)}$.)
{\it Along all trajectories that do restore split-symmetry,  $\cfunc_k$ was found to be perfectly monotone,} and stationary both at the NGFP and in the classical regime.
Unphysical trajectories, not restoring split-symmetry in the IR, on the other hand, can give rise to a non-monotone behavior of $\cfunc_k$.

\begin{figure}[!ht]
\centering
\centering
\psfrag{b}[cm][0][1][90]{${\scriptscriptstyle k\partial_k  1\slash \cF_k}$}
\psfrag{a}{${\scriptscriptstyle k \slash m_{\text{Pl}}}$}
\psfrag{k}{${\scriptstyle k \slash m_{\text{Pl}}}$}
\psfrag{x}[r]{${\scriptstyle 1\slash \cF_k }$}        
 \subfloat{
 \psfrag{c}[cm][0][1][90]{${\scriptstyle 1\slash \cF_k }$}  
 \includegraphics[width=0.450\textwidth]{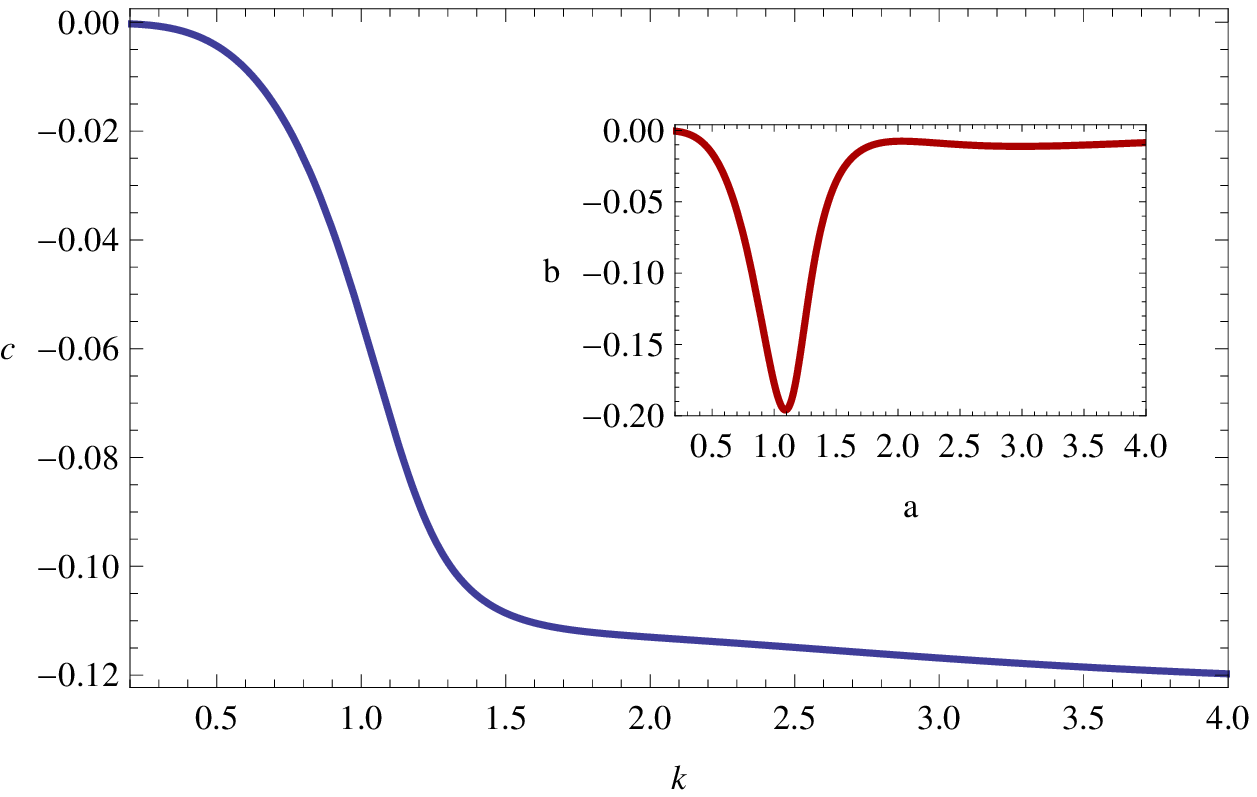}
 } 
 \hspace{0.04\textwidth} %
 \subfloat{%
 \psfrag{b}[cm][0][1][90]{${\scriptscriptstyle k\partial_k 1\slash \cF^{\sm}_k}$}
 \psfrag{c}[cm][0][1][90]{${\scriptstyle 1\slash \cF^{\sm}_k }$}  
\includegraphics[width=0.455\textwidth]{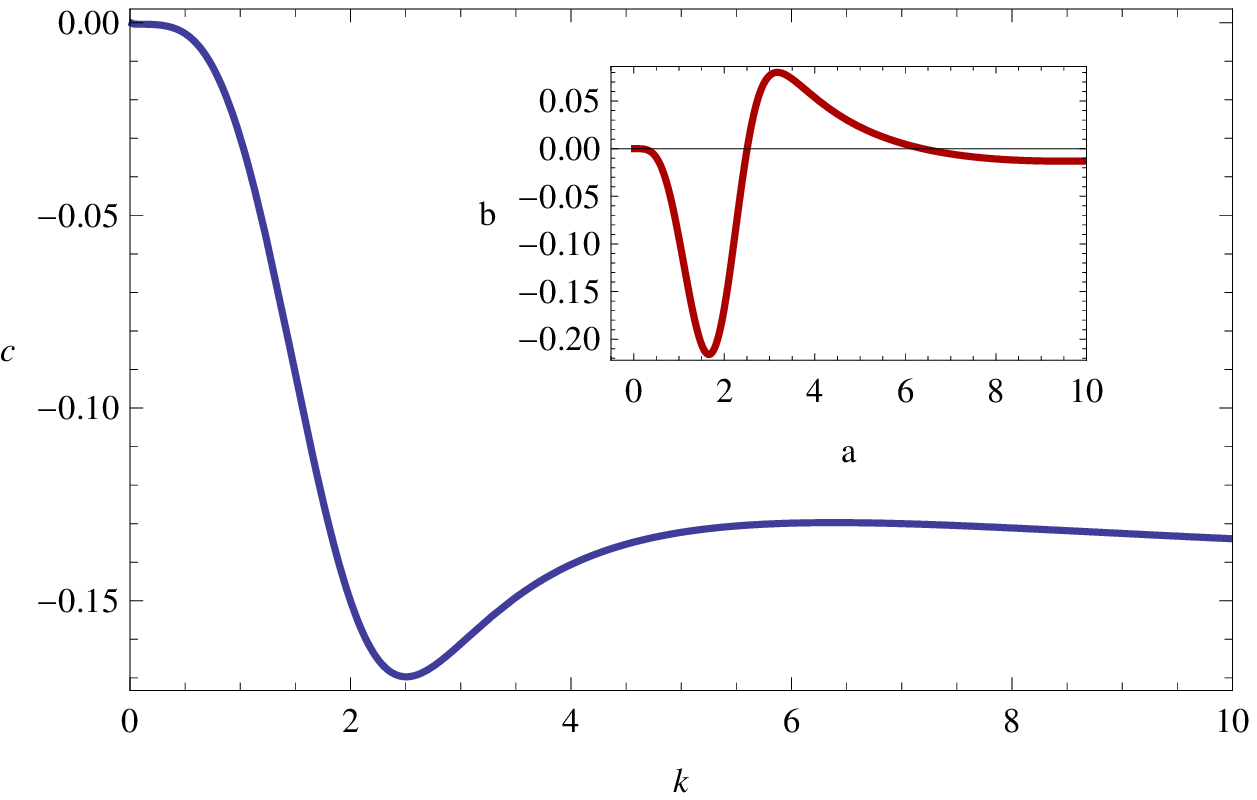}
}
\caption{The function $1\slash \cF_k$ computed from the bi-metric (left diagram) and the single-metric truncation (right diagram), respectively, along a representative type \Rmnum{3}a trajectory.
The monotonicity of $\cfunc_k$ is violated if the scale derivative of $1\slash \cF_k$, shown in the two insets, assumes positive values.} \label{fig:cfuncBaSM}
\end{figure}

A similar analysis was performed on the basis of the single-metric version of the Einstein-Hilbert truncation with a 2 dimensional theory space.
It is less precise than its bi-metric counterpart as it hypothesizes perfect split-symmetry on all scales, something that can be true at best approximately because of the various unavoidable sources of symmetry breaking in the EAA  (cutoff action $\Delta S_k$, gauge fixing term).
Regarding the monotonicity of $\cfunc_k$, we found that {\it $\cfunc_k$ fails to be monotone for any of the single-metric type \Rmnum{3}a trajectories.}
The detailed analysis revealed that this failure is due to the (not quite unexpected) insufficiency of the single-metric approximation, rather than to a structural defect of the candidate $\cfunc_k=\Gamma_k[0;\bar{\Phi}^{\scon}_k]$.
For a typical RG trajectory, both the single- and bi-metric $\cfunc_k$-functions are depicted in Fig. \ref{fig:cfuncBaSM}.

The numerical results \cite{daniel-C} lend strong support to the following 

{\noindent\bf Conjecture:} 
{\it In the full theory, QEG in 4 dimensions, or in a sufficiently general truncation thereof, the proposed candidate for a `$C$-like' function is a monotonically increasing function of $k$ along all RG trajectories that restore split-symmetry in the IR and thus comply with the fundamental requirement of Background Independence.}

{\noindent\bf Crossover trajectories and entropy of de Sitter space.}
The function $\cfunc_k$ for truncated QEG is stationary  at fixed points as well as in classical regimes.
This is obvious from the following two alternative representations of the reduced $\cfunc_k$-function:
\begin{align}
\cF_k=-\frac{2\Kk_k^{(1)}-\Kk_k^{(0)}}{\tg_k^{(0)}\big(\Kk_k^{(1)}\big)^2}=-\frac{2\Kkbar^{(1)}_k-\Kkbar_k^{(0)}}{G_k^{(0)}\big(\Kkbar_k^{(1)}\big)^2}
\end{align}
We see that $\cF_k$, and hence $\cfunc_k$ becomes stationary when the {\it dimensionless} couplings are at a fixed point of the flow, and also when the {\it dimensionful} ones become scale independent; this is the case in a classical regime (`$\crg$') where, by definition, no physical RG effects occur.
If $\Kkbar_{\crg}^{\cix}$ and $G_{\crg}^{\cix}$ denote the constant values of the cosmological and Newton constants there, this regime amounts to the trivial canonical scaling $\Kk_k^{\cix}=k^{-2}\Kkbar_{\crg}^{\cix}$ and $\tg_k^{\cix}=k^{d-2}G_{\crg}^{\cix}$.

As a result, there exists the possibility of generalized crossover transitions, not  in the standard way from one fixed point to another, but rather from a fixed point to a classical regime or vice versa.
Thereby $\cfunc_k$ will always approach well defined stationary values $\cfunc_*$ and $\cfunc_{\crg}$ in the respective fixed point or classical regime.
In quantum gravity, the investigations of such generalized crossover transitions is particularly important since one of its main tasks consists in explaining the emergence of a classical spacetime from the quantum regime.

Specializing again for an asymptotically safe type \Rmnum{3}a trajectory, the initial point in the UV is a non-Gaussian fixed point.
For the limit $\cfunc^{\UV}\equiv \lim_{k\rightarrow\infty}\cfunc_k$ the bi-metric calculation yields $\cfunc^{\UV}=\cfunc_*$, with
\begin{align}
\cfunc_*=-\frac{2 \Kk_*^{(1)}-\Kk_*^{(0)} }{\tg_*^{(0)}\big(\Kk_*^{(1)} \big)^2}\, \mathcal{V}(\MaFs,\gInst)
\label{eqn:cm_cross_01}
\end{align}
According to the Einstein-Hilbert results for the NGFP, $-\cfunc_*\slash \mathcal{V}(\mathcal{M},\gInst)$ is a positive number of order unity, presumably between about $4$ and $8$.
Concerning the opposite limit $\cfunc^{\IR}=\lim_{k\rightarrow0} \cfunc_k$,  the  trajectory describes a generalized crossover, enters  a classical regime, and restores split-symmetry for $k\rightarrow 0$.
This entails  $\cfunc^{\IR}=\cfunc_{\crg}$ where
\begin{align}
\cfunc_{\crg} =-\frac{ \mathcal{V}(\MaFs,\gInst) }{G_{\crg}\Kkbar_{\crg} } 
\label{eqn:cm_cross_03}
\end{align}
Here we exploited that split-symmetry implies the values of $G_{\crg}$ and $\Kkbar_{\crg}$ to be level independent.

We may conclude that {\it in  an asymptotically safe theory of quantum gravity which is built upon a generalized crossover trajectory from criticality (the {\rm NGFP}) to classicality  the $\cfunc_k$-function candidate implies the `integrated $\cfunc$-theorem'  $\nBound\equiv \nBound_{0,\infty}=\cfunc^{\UV}-\cfunc^{\IR}$ with finite numbers $\cfunc^{\UV}$, $\cfunc^{\IR}$, and $\nBound$}. 

This finiteness is in marked contrast to what standard perturbative field theory would predict. 
Clearly, the Asymptotic Safety of QEG is the essential prerequisite for this property since it is the non-Gaussian fixed point that  assigns a well defined, computable value to $\cfunc^{\UV}$.

The quantity $\nBound$ can be interpreted as a measure for the `number of modes' which are integrated out while the cutoff is lowered from infinity to $k=0$.
The notion of `counting' and the precise meaning of a `number of field modes' is defined by the EAA itself, namely via the identification $\cfunc_k=\EAA_k[\bar{\Phi}^{\scon}_k,\bar{\Phi}^{\scon}_k]$.
We saw in the previous section that, under special conditions, $\cfunc_k$ is literally counting  the $\EAA_k^{(2)}$-eigenvalues in a given interval.
However, generically we are dealing with a non-trivial generalization thereof which, strictly speaking, amounts to a {\it definition} of `counting'.
As such it is  probably the most natural one from the perspective of the EAA and the geometry of theory space.

Let us consider a simple caricature of the real Universe, namely a family of de Sitter spaces along a type \Rmnum{3}a trajectory, whose classical regime in the IR has $\Kkbar_{\crg}>0$.
Assuming it represents the real final state of the evolution, we have $\cfunc^{\IR}=-3\pi\slash G_{\crg}\Kkbar_{\crg}<0$.
Note that $|\cfunc^{\IR}|$ equals precisely the well known semi-classical Bekenstein-Hawking entropy of de Sitter space.

If in particular $G_{\crg} \Kkbar_{\crg}\ll 1$, corresponding to a very `large' classical Universe, we have $|\cfunc^{\IR}|\gg 1$, while $|\cfunc^{\UV}|=\Order{1}$ is invariably determined by the NGFP coordinates.
As a consequence, the number $\nBound$ is completely dominated by the IR part of the trajectory:
\begin{align}
\nBound= \cfunc^{\UV}-\cfunc^{\IR}\approx - \cfunc^{\IR}\approx + \frac{3\pi}{G_{\crg} \Kkbar_{\crg}}\gg 1
\label{eqn:cm_cross_07}
\end{align}
Identifying $\Kkbar_{\crg}$ and $G_{\crg}$ with the corresponding values measured in the real Universe we would find $\nBound\approx 10^{120}$.

Thus, in the sense explained above, the familiar Bekenstein-Hawking entropy of de Sitter space acquires a rather concrete interpretation, namely as  the number of metric and ghost fluctuation modes that are integrated out between the NGFP in the UV and the classical regime in the IR.
It plays a role analogous to the central charge of the IR conformal field theory in Zamolodchikov's case.

Concerning the finiteness of $\nBound$, the situation changes fundamentally if we try to define the function $k\mapsto \cfunc_k$ along trajectories of the type \Rmnum{1}a, those heading for  a negative cosmological constant $\KkD$ after leaving the NGFP regime, and of type \Rmnum{2}a, the single trajectory which crosses over from the NGFP to the Gaussian fixed point.
For all {type \Rmnum{1}a} trajectories, $\cfunc_k$ becomes singular at some nonzero scale $k_{\text{sing}}>0$ when they pass  $\KkD=0.$ 
As eq. \eqref{eqn:intro_eq11} shows, $\cF(\,\cdot\,)$ and $\cfunc(\,\cdot\,)$ have a pole there so that $\cfunc_k$ diverges in the limit $k\searrow k_{\text{sing}}$.
The number  $\nBound_{k_{\text{sing}},\infty}$ is infinite then, even though not all modes are integrated out yet.
There is a non-trivial RG evolution also between $k_{\text{sing}}$ and $k=0$.
The tadpole equation has qualitatively different solutions for $k>k_{\text{sing}}$, $k=k_{\text{sing}}$, and $k<k_{\text{sing}}$, namely spherical, flat, and hyperbolic spaces, respectively ($S^d$, $R^d$, and $H^d$, say).
This topology change prevents us from smoothly continuing the mode count across the $\KkD=0$ plane.
This is the reason why we mostly focused on type \Rmnum{3}a trajectories here. 

\section{Summary}\label{sec:conc}
The effective average action is a variant of the standard effective action which has an IR cutoff built in at a sliding scale $k$.
At least for systems without fermions it possesses a natural mode counting and (`pointwise') monotonicity property which is strongly reminiscent of, but actually not equivalent to, Zamolodchikov's $C$-function in 2 dimensions.
Motivated by this observation, and taking advantage of the structures and tools that are naturally provided by the manifestly non-perturbative EAA framework, we tried to find a map from the functional $\EAA_k[\Phi,\bar{\Phi}]$ to a single real valued function $\cfunc_k$ that shares two main properties with the $C$-function in 2 dimensions, namely monotonicity along RG trajectories and stationarity at RG fixed points.
Such a map is unlikely to exist in full generality.
In fact, an essential part of the research program we are proposing consists in finding suitable restrictions on, or specializations of the {\it admissible trajectories} (restoring split-, or other symmetries, etc.), the {\it theory space} (with respect to field contents and symmetries), the underlying {\it space of fields} (boundary conditions, regularity requirements, etc.), and the {\it coarse graining methodology} (choice of cutoff, treatment of gauge modes, etc.) that will guarantee its existence.
We motivated  a specific candidate for a map of this kind, namely $\cfunc_k=\EAA_k[\bar{\Phi}^{\scon}_k,\bar{\Phi}^{\scon}_k]$ where $\bar{\Phi}^{\scon}_k$ is a running self-consistent background, a solution to the tadpole equation implied by $\EAA_k$.
This function $\cfunc_k$ is stationary at fixed points, and a non-decreasing function of $k$ provided the breaking of the split-symmetry which relates fluctuation fields and backgrounds is sufficiently weak.
Thus, for a concrete system the task is to identify the precise conditions under which the split-symmetry violation does not destroy the monotonicity property of $\cfunc_k$, and to give a corresponding proof then. 

By means of a particularly relevant example, asymptotically safe QEG in 4 dimensions, we demonstrated that this strategy is viable in principle and can indeed lead to interesting candidates for `$C$-like' functions under conditions which are not covered by the known $c$- and $a$-theorems.
Within a sufficiently precise truncation of QEG, on a 4 dimensional theory space, we showed that $\cfunc_k$ has exactly the desired properties of monotonicity and stationarity, provided it is based upon a  physically meaningful RG trajectory, that is, one which leads to a restoration of Background-Independence once all field modes are integrated out.

\subsection*{Acknowledgment}
\noindent M.~R. would like to thank the organizers of {\it Quantum Mathematical Physics} for their hospitality at Regensburg and for a particularly stimulating conference.

\end{document}